\documentclass[twocolumn]{svjour3}          
\usepackage{graphicx}
\begin{document}

\title{Reentrant superconductivity in YBa$_2$Cu$_3$O$ _{7-\delta}$  microstructured particles
}

\author{Rodolfo E. L\'{o}pez-Romero* \and Dulce Y. Medina \and R. Escudero  
}

\institute{Rodolfo E. L\'{o}pez-Romero \at
              Divisi\'{o}n de Ciencias B\'{a}sicas e Ingenier\'{i}a, Universidad Aut\'{o}noma Metropolitana-Azcapotzalco, Av. San Pablo No 180, Col. Reynosa-Tamaulipas, C.P. 02200 M\'{e}xico D.F., M\'{e}xico.  \\
              Tel.: (5255) 5622-4625\\
              \email{rol\_eze@ciencias.unam.mx}           
           \and
           Dulce Y. Medina \at
              Divisi\'{o}n de Ciencias B\'{a}sicas e Ingenier\'{i}a, Universidad Aut\'{o}noma Metropolitana-Azcapotzalco, Av. San Pablo No 180, Col. Reynosa-Tamaulipas, C.P. 02200 M\'{e}xico D.F., M\'{e}xico.
               \and
              R. Escudero \at
              Instituto de Investigaciones en Materiales, Universidad Nacional Aut\'{o}noma de M\'{e}xico. A. Postal 70-360. M\'{e}xico, D.F.
}

\date{Received: date / Accepted: date}

\maketitle

\begin{abstract}
		
A reentrant superconducting effect was observed in YBCO particles as a consequence of their size and microstructure. It rises from the coexistence between bulk superconductivity and surface ferromagnetism,  where the enhanced magneto-electric coupling induced by broken spatial-inversion symmetry in layers near structural defects plays a very important role on surface ferromagnetism. The study assumes grain boundaries inside the particles act as additional surface regions that enhance the ferromagnetic properties. Magnetic and superconducting properties as a function of particle size were studied. Particularly, the 385 nm particles which were the ones that exhibited the reentrant superconducting effect. The Ginzburg-Landau parameter and the critical temperatures are \textit{k} $\sim $ 1.29 $\pm$.03,  T$ _{c1} $ = 92 $\pm$0.5 K, and  T$ _{c2} $ $ \approx $ 10 $\pm$3 K.


\end{abstract}

\section{Introduction}
\label{intro}
Reentrant superconductivity is the most interesting feature observed in Kondo and ferromagnetic superconductors. The nature of this phenomenon is attributed to the presence of magnetic ions in a superconducting host, hence they can exhibit, or not, a long-range ferromagnetic order \cite{Rogalla} \cite{Riblet} \cite{Fertig} and \cite{Ishikawa}. Reentrant superconductivity was observed for the first time in the Kondo system (La$ _{1-x} $Ce$ _{x} $)Al$ _{2} $ and later on the Ferromagnetic Superconductors (FS) ErRh$ _{4} $B$ _{4} $ and Ho$ _{1.2} $Mo$ _{6} $S$ _{8} $ \cite{Fertig} and \cite{Ishikawa}. These compounds become superconducting at a critical temperature T$ _{c1} $, and  return to the normal state at a second critical temperature T$ _{c2} $. The return to the normal state coincides with the occurrence of the Kondo effect or long-range ferromagnetic order. Particularly, FS systems are interesting because they could help us understand the close relationship between superconductivity and magnetism.

Recently, reentrant superconductivity has also been studied in systems where superconductivity and ferromagnetism exist in physically delimited regions, for example, in artificially layered Ferromagnet/Superconductor (F/S) structures. The most spectacular result in these systems is the oscillatory behavior of T$ _{c} $ as a function of ferromagnetic layer thickness d$ _{F} $ \cite{Zdravkov}.

On the other hand, much attention has been paid in nano-superconductors, including not only conventional superconductors \cite{Reich} \cite{HLi} and \cite{WHLi} but also those with high-T$ _{c} $ \cite{Yuewei} \cite{Shipra} \cite{Hasanain} and \cite{Zhonghua}. The size alters both magnetic and electronic properties. It is well known that bulk superconductors with a high-concentration of microcracks, voids, dislocations, and grain-boundaries constructed by random stacks of small grains show a suppression of the super-current \cite{Stoneham}. However, recent studies in high-T$_{c}$ cuprates nanoparticles grown in a non-chemical equilibrium state present weak ferromagnetism at room temperature, which gives new insights into the fundamental properties of the superconducting state. Shipra and co-workers \cite{Shipra} and Sundaresan et al. \cite{ASundaresan} and \cite{Sunda} have reported this behavior in YBCO nanoparticles. Hasanain et al. \cite{Hasanain}  have investigated the particle size dependence in the magnetic and superconducting properties, studying the relationship between particle size and the presence of a long-range ferromagnetic order inside the superconducting state. They concluded a possible coexistence between surface ferromagnetism and bulk superconductivity. Additionally, Zhu et al. \cite{Zhonghua} and Fan et al. \cite{Fan} showed that ferromagnetism in YBCO nanoparticles may be associated with either; surface oxygen vacancies or structural defects. In the former case, surface oxygen vacancies lead to a charge redistribution between different surface ions \cite{Coey}, while in the second case mimicking small YBCO clusters with different surfaces using \textit{ab-initio} Density Functional Theory (DFT) is shown that the structural defects lead to an enhancement of the magneto-electric coupling due to the broken spatial-inversion symmetry in layers near structural defects. In both cases there are many unsatisfied bonds, so a large number of uncompensated spins can be coupled via direct or indirect leading to the ferromagnetic order \cite{HLi}. Some research based on theoretical models and \textit{ab-initio} (DFT) simulations show that the partial occupation of \textit{p} orbitals of the oxygen atoms near structural defects induces ferromagnetism in compounds that, in principle, are not magnetic \cite{Fan} \cite{Osorio} and \cite{Chaitanya}. In fact, a surface is a two-dimensional structural defect, and atoms near the surface have a lower coordination number than those in bulk \cite{Fan} and \cite{Chaitanya}.
   
It is important to mention that the low coordination number narrows the electronic bands of YBCO surface states, as well as increases the local density of states (DOSs) near the Fermi energy, which in consequence induces the magnetic instability of surface electrons, forming the ferromagnetic order by electron-electron exchange interaction \cite{Fan} and \cite{Junling}. This means that the electric field produced by broken spatial-inversion symmetry can induce the spin-polarization of surface electrons even without oxygen vacancies, giving place to the surface ferromagnetism \cite{Fan}.

In order to study the magnetic effects induced by particle size reduction and microstructure in a high-T$ _{c} $ superconductor, YBCO microstructured particles at different sizes were synthesized. The study brings to the foreground three main aspects: First, it is considered that the source of the ferromagnetic properties in the superconducting system is attributed to the low coordination number of the surface atoms, a well-known fact \cite{Fan} \cite{Osorio} and \cite{Chaitanya}. Second, grain boundaries inside the particles are suggested to act as additional surface regions that enhance the ferromagnetic properties. And third, the enhanced ferromagnetic properties are proposed as the main factor that takes the particles out of the superconducting state at some temperature below T$ _{c} $. Our results show that the increase of surface/volume ratio in the particles weakens the superconductivity and enhances the ferromagnetism simultaneously.

\section{Materials and Methods}
\label{sec:1}

YBCO microstructured particles were obtained by a mechanical grinding process, and the bulk sample was synthesized by solid-state reaction. Yttrium Oxide (6N), Barium Carbonate (5N), and Copper Oxide (5N) in a stoichiometry 1:4:6 were used. Reagents were mixed and calcined at 950 $^{\circ}$C for 24 h. Subsequently, the mixture was pelletized and sintered two times in a flowing oxygen atmosphere at 950 $^{\circ}$C for 24 h with a cooling rate of 50 $^{\circ}$C/h \cite{Jin}. Finally, the YBCO bulk sample was ground in an agate mortar, and the resulting powder was suspended in pure acetone in order to classify the particle size as a function of precipitation time. In this case, the mechanical grinding process was used not only to break the bulk superconductor but also to induce microstructure in the particles. The presence of structural defects such as dislocations, grain boundaries, and micro-strains tends to decrease the crystal size in the particles. Ferromagnetism in nanoparticles is essentially a surface phenomenon associated with the enhanced defect concentration at the surface. If it is assumed that grain boundaries act as additional surface regions, then an enhanced of the ferromagnetic properties will occur in the particles with the highest surface/volume ratio. Using this synthesis procedure the ferromagnetism associated with surface oxygen vacancies is despised because they are greatly reduced by atomic diffusion processes, a fact that it does not happen in other synthesis procedures (Citrate pyrolysis \cite{Hasanain} and \cite{Zhonghua}, Citrate-Gel \cite{Blinov} and Co-precipitation \cite{AManthiram}).

\begin{figure}

	\includegraphics[width=0.5\textwidth]{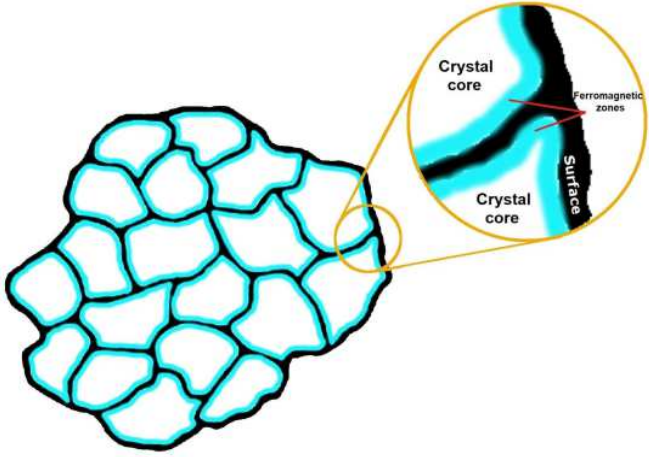}
	\caption{(Color on-line) Schematic representation of a microstructured particle. Grain boundaries are suggested as particle surface area. The enhanced magneto-electric coupling induced by broken spatial-inversion symmetry in layers near structural defects can induce the spin-polarization by electron-electron exchange interaction \cite{Fan}.}
	\label{fig:1}       
\end{figure}

\section{Results and discussion}

The study basically considers that grain boundaries inside the particles act as additional surface regions that enhance the ferromagnetic properties, therefore a microstructured particle is treated as a polycrystalline system formed by a set of crystals, Fig. 1. The surface/volume ratio ($SA:V$) was evaluated as a function of particle size. Particle size was determined by Dynamic Light Scattering (DLS) measurements, and complemented with Transmission Electron Microscopy (TEM), Fig.2(\textit{a-d}), while crystal size was evaluated by Rietveld refinement of X-ray diffraction data, Fig.2\textit{e}.

YBCO phase crystallizes in $Pmmm$ orthorhombic system, with lattice parameters $a$=3.817(9), $b$=3.880(1) and $c$=11.665(5) nm (COD ID 100-1435). Bulk YBCO was synthesized following the normal known conditions \cite{Jin}, so the compound does not present oxygen deficit (or vacancies), aspect different as reported by Zhonghua et al. \cite{Zhonghua}. 

Usually, a polycrystalline sample presents a high density of structural defects such as dislocations, grain boundaries, and micro-strains, which manifest in the X-ray diffraction data by a broadening of the Bragg peaks. Through this data, it is possible to evaluate the crystal size in the sample. According to the above, in Fig. 3\textit{a} is shown the particle and crystal size as a function of precipitation time. Crystal size in YBCO bulk was estimated around 52 nm and decreases with the particle size. In Fig. 3\textit{b} is shown the experimental ratio between particle and crystal size, \textit{R} is a parameter related to the average number of crystals that form the particles.

In theory, a polycrystalline particle can be treated as a set of crystals, where each crystal occupies the same volume as a sphere with diameter $ d_ {c} $. Then the surface area and volume of a crystal are given by:

\begin{equation}
	A_{c}=\pi d_{c}^{2};
\end{equation}
\begin{equation}
	V_{c}=\frac{\pi d_{c}^{3}}{6}
\end{equation}

so, the $ SA:V $ ratio for a particle is given by:

\begin{equation}
	SA:V =\frac{SA }{V_{p}}
\end{equation}

where $SA $ is the particle surface area, given by $ A_{c}*R $, and $ V_{p} $ is the volume of the particle, given by $ \pi d_{p}^{3}/6$ ($ d_{p} $ is equal to the particle diameter).

In Fig. 3\textit{c} is shown the fit of the Crystal size \textit{vs.} Particle size experimental curve. From this data was deduced the $d_{c}$ parameter.

\begin{figure*}
	\centering
	\includegraphics[width=0.9\textwidth]{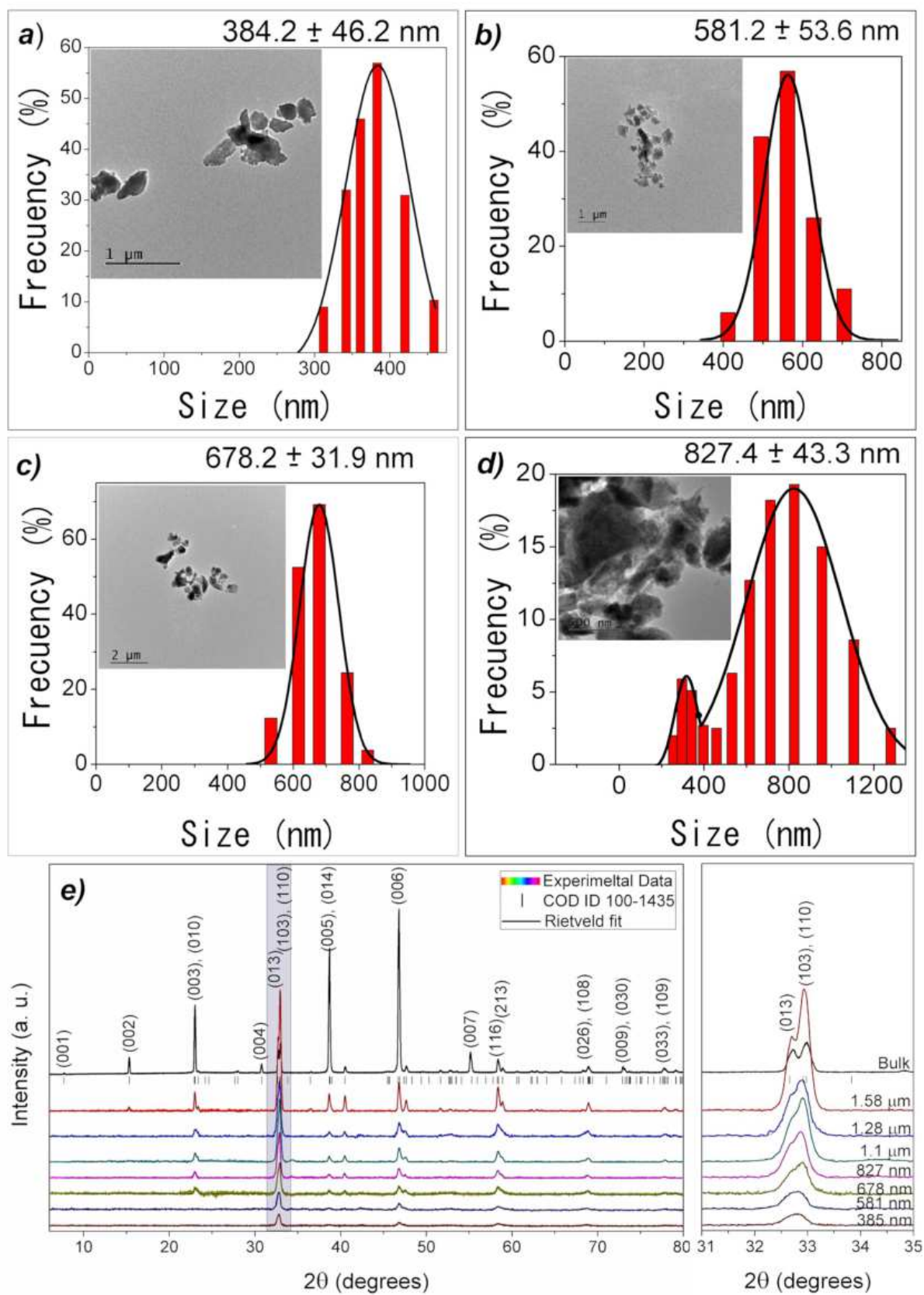}
	\caption{(Color on-line) Particle size distributions (\textit{a-d}). X-ray diffraction data and Rietveld refinement (\textit{e}). An enlargement of the gray area in \textit{e} is shown in the  right inset. COD ID 100-1435 card corresponds to the YBa$_{2}$Cu$ _{3} $O$ _{6.92} $ phase.}
	\label{fig:2}       
\end{figure*}

Two cases can be deduced from eq. (3); one in which a crystal is equal to a particle, monocrystalline regime, and other in which a particle is a set of crystals, polycrystalline regime. 

Fig. 3\textit{d} shows a comparison between the normalized $SA:V$ experimental and theoretical ratio. Both regimes, monocrystalline as well as polycrystalline, are presented. The dotted red line represents the monocrystalline regime where $ SA $ in eq. (3) is reduced to $ \pi d_{p}^{2} $. In this case, an exponential behavior of $SA:V$ ratio is observed when particle size decreases, implying an increase of the particle surface area, a well-known fact \cite{Hasanain}. In contrast, a remarkable increase of surface/volume ratio is observed when grain boundaries are considered as particle surface area (dotted blue line). Regardless of the line-like behavior exhibited by $SA:V_{poly} $, such increment can be interpreted as an enhancement of the ferromagnetic properties since this behavior is directly linked to the structural defects. Particles with a high density of structural defects will be highly influenced by ferromagnetic order.

\begin{figure*}
	\centering
	\includegraphics[width=1\textwidth]{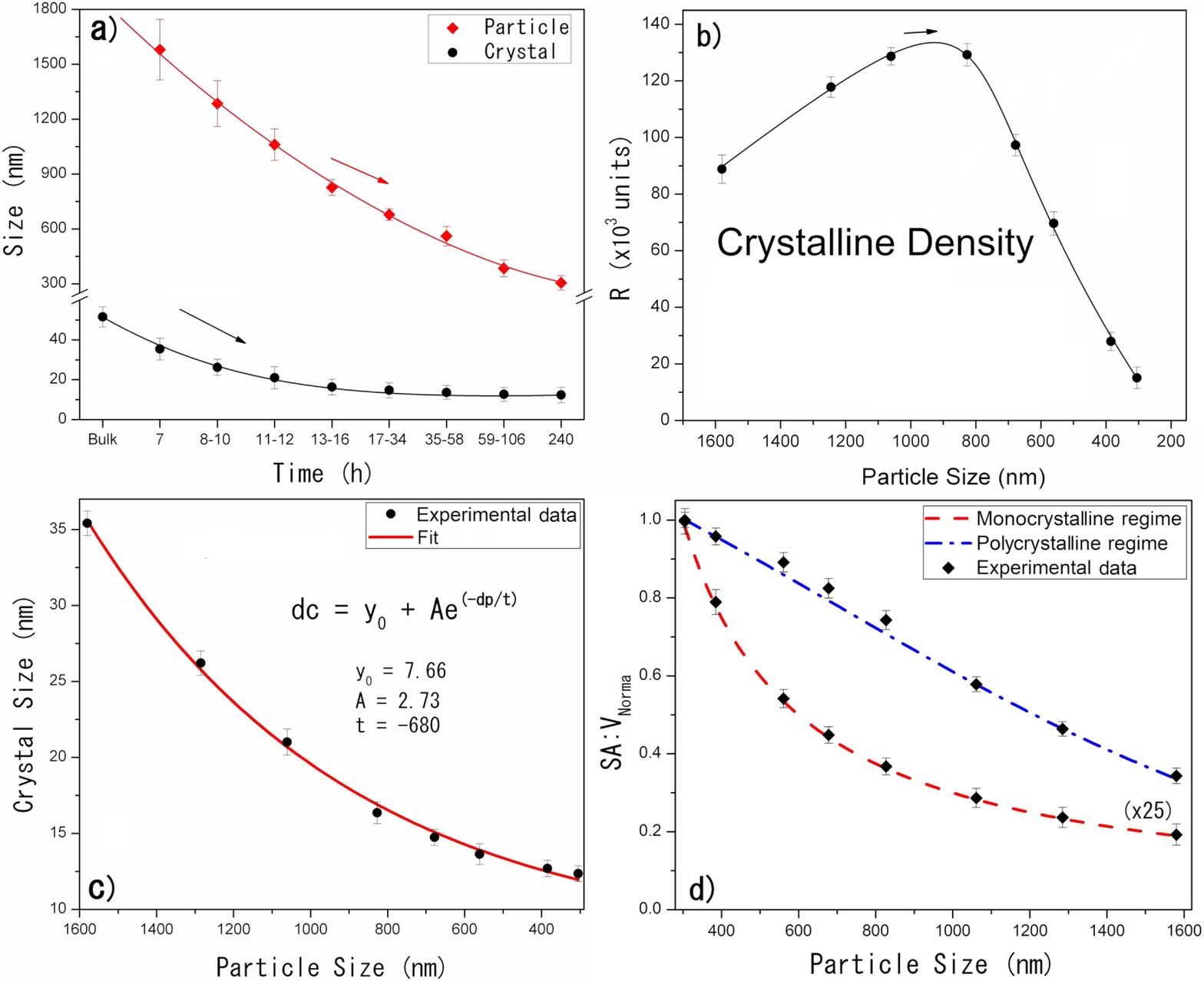}
	\caption{(Color on-line) Particle and crystal size as a function of precipitation time \textit{a)}. Crystalline density as a function of particle size \textit{b)}. Crystal size as a function of particle size \textit{c)}. The dotted red line represents the theoretical fit through the $d _{c} $ function. Normalized surface/volume ratio in both regimes is shown in \textit{d)}. Monocrystalline regime only considers the particle surface area, while the polycrystalline considers both particle surface and grain boundaries. Dotted lines represent the theoretical fits according to eq. (3) (based on the approach crystals taken as spheres) while black dots are the experimental data (Fig. 3\textit{a}). The dotted red line was multiplied by 25 for more clarity.}
	\label{fig:3}       
\end{figure*}

\begin{figure}
	\includegraphics[width=.49\textwidth]{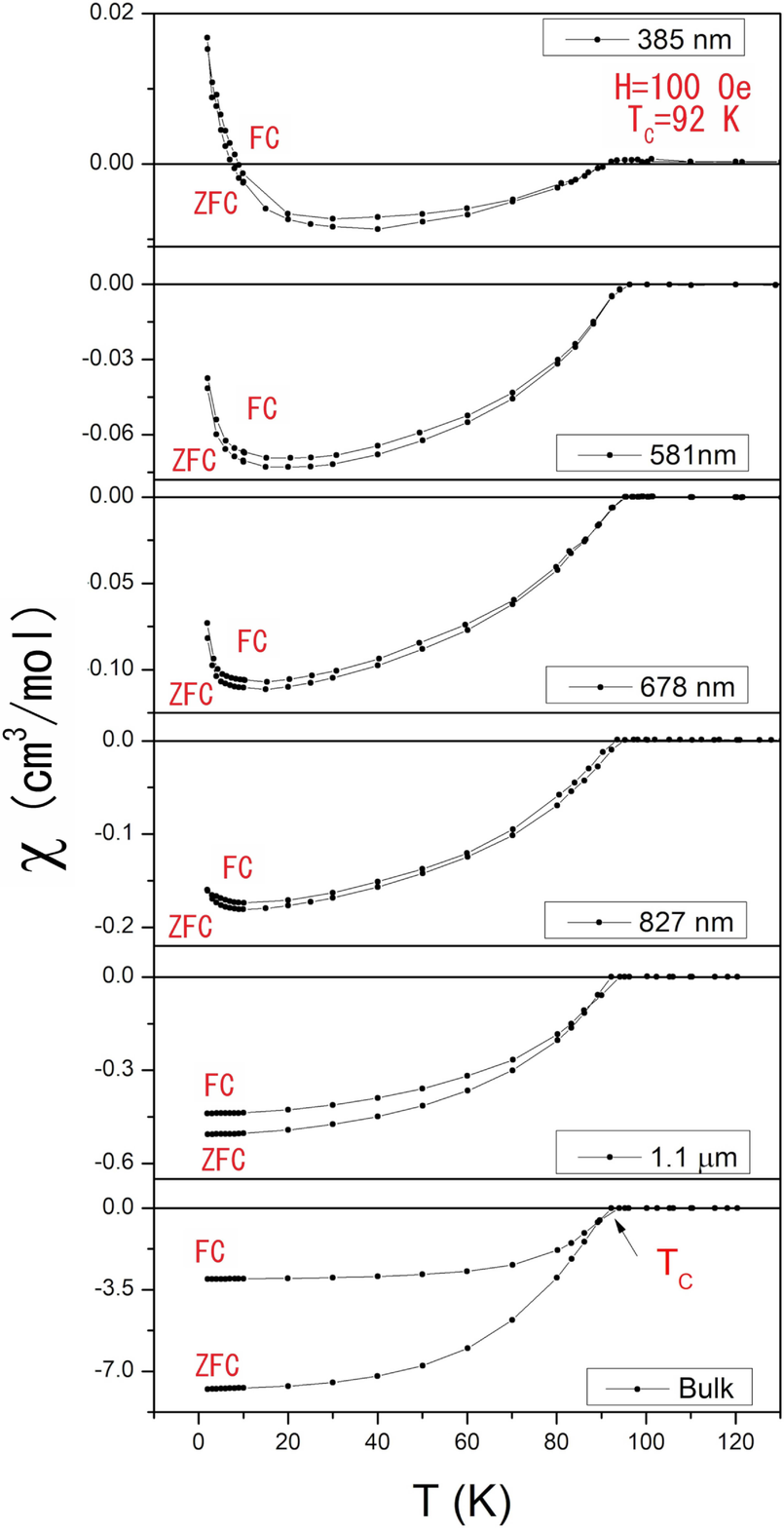}
	\caption{(Color on-line) $M(T)$ measurements in YBCO microstructured particles at different sizes. ZFC and FC correspond to the Zero Field Cooling and Field Cooling conditions, respectively. The reentrant superconducting effect is observed at $\approx$ 10 K in the 385 nm particles. The paramagnetic contribution was subtracted from raw data according to Gotor et al. in particles less than 678 nm \cite{J.Gotor}.}
	\label{fig:4}       
\end{figure}

In Fig. 4 is shown magnetic susceptibility measurements  ($\chi(T)$) in YBCO bulk as well as in microstructured particles. The bulk sample shows a critical temperature T$ _{c} \sim $ 92 K. However, when particle size decreases $\sim $ 830 nm, the magnetization at low temperatures shows a successive increase, which even in the 385 nm particles this characteristic becomes non-diamagnetic, showing the transition to the normal state. This behavior has already been reported in Kondo and ferromagnetic superconductors \cite{Riblet} \cite{Fertig} \cite{Ishikawa} and \cite{Maple}. Nevertheless, this is the first experiment in which reentrant superconductivity is observed in a high-$T_{c}$ superconductor as a consequence of its size and microstructure. Reentrant superconductivity is characterized by a return transition from the superconducting to the normal state at a lower temperature than T$ _{c} $. The nature of this phenomenon is attributed to the presence of magnetic ions in a superconducting host. The electromagnetic interaction between magnetic ions and superconducting electrons gives place to the breaking of Cooper pairs, suppressing the superconductivity. Nonetheless, in the present study, it is proposed that the reentrant superconducting effect is caused by the coexistence between bulk superconductivity and surface ferromagnetism where layers near structural defects enhance the ferromagnetic order while crystal cores remain superconducting.

\begin{figure*}
	\centering
	\includegraphics[width=1\textwidth]{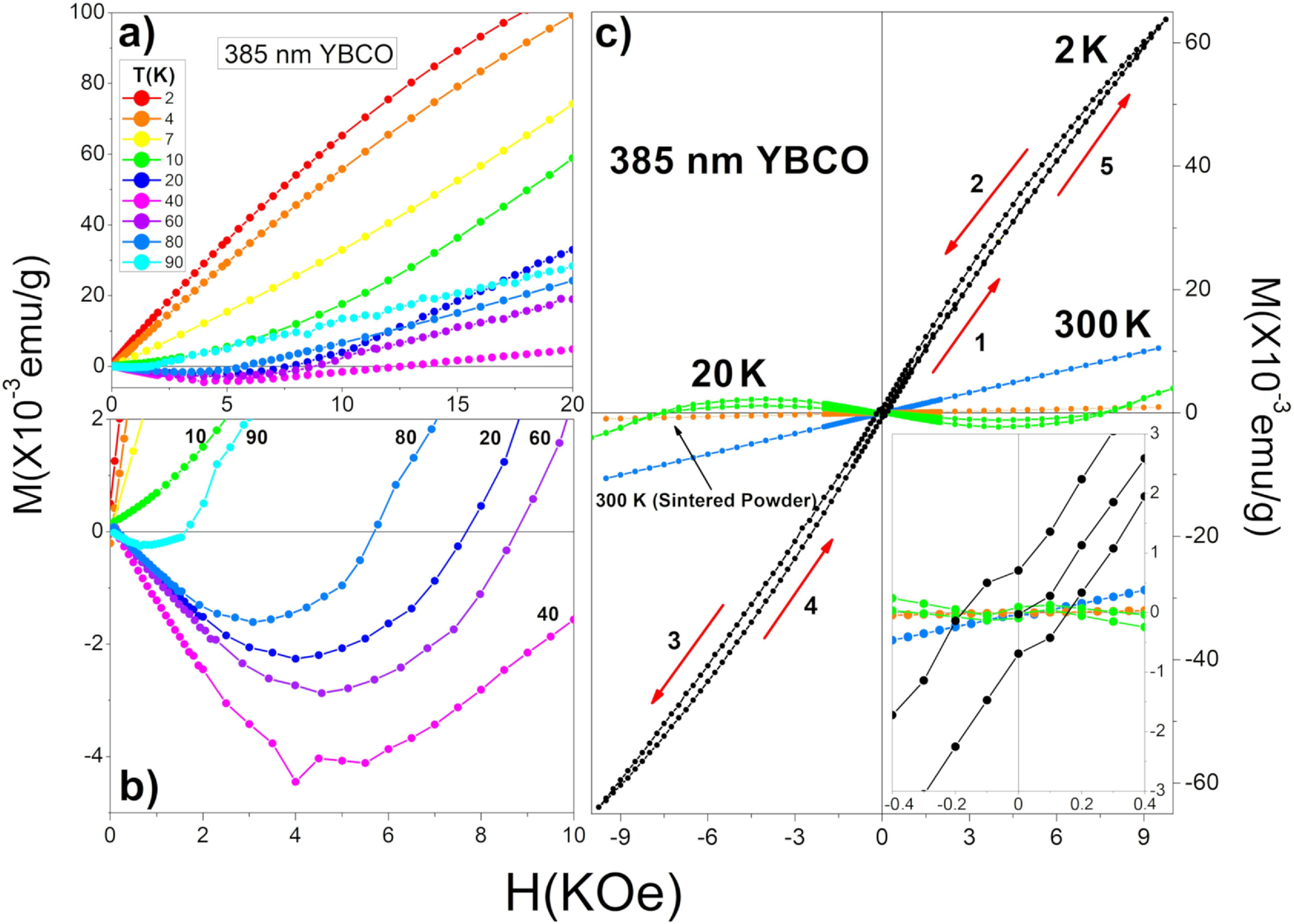}
	\caption{(Color on-line) \textit{M(H)} measurements as a function of temperature in the 385 nm YBCO particles a), b). Isotherms at 2, 20, and 300 K are shown in c). Ferromagnetism is observed at 2 K with a small coercive field. Inset shows the hysteresis loops at low-applied magnetic fields. Both 385 nm particles and sintered powder show paramagnetism at 300 K.}
	\label{fig:5}       
\end{figure*}

Diverse theoretical studies based on Landau's theory of second-order phase transitions have shown that the enhanced magneto-electric coupling induced by broken spatial-inversion symmetry in layers near structural defects plays a very important role on surface ferromagnetism. Atoms near particle surface have lower coordination number than those in bulk \cite{Fan} and \cite{Junling}. The low coordination number involves a narrowing of the electronic bands of YBCO surface states, while the electric neutral condition induced by local broken of spatial-inversion symmetry promotes the charge redistribution \cite{Fan} and \cite{Coey}. Surface ferromagnetism is similar to an itinerant magnetism associated with a high-correlated electron gas on the surface. The high density of states at the Fermi level determines the Stoner splitting condition to develop spontaneous spin-polarization \cite{Yuewei}. Surface ferromagnetism above room temperature has been previously reported in YBCO nanoparticles and although it is associated with the presence of surface oxygen vacancies it continues to be intriguing \cite{Shipra} \cite{Hasanain} \cite{Zhonghua} \cite{ASundaresan} and  \cite{Sunda}. It is known that surface oxygen vacancies facilitate the charge transfer and bring the Fermi level up to a peak in the local density of states, leading to Stoner splitting of \textit{N}$_{s}  $(\textit{E}) \cite{Hasanain} \cite{Zhonghua} \cite{Osorio} and \cite{Chaitanya}. In the absence of them or any other cation, the electric neutral condition induced by the local broken of spatial-inversion symmetry promotes the charge redistribution and electrons in the bulk can redistribute near particle surface under the interaction with the internal electric field \cite{Fan}. Certainly, although it is known that a large number of uncompensated spins can be expected not only from the particle surface but also from the grain boundaries, the exchange-correlation interaction between the surface electrons may be the liable condition for the low ferromagnetic temperature observed in the present study. This could happen only if the Fermi level lies far from the peak in \textit{N}$_{s}  $(\textit{E}) \cite{HLi} \cite{Coey} and \cite{Wahle}.

At low temperatures and small particle sizes, the electromagnetic interaction between bulk superconductivity and surface ferromagnetism tends to suppress the superconductivity, observing the transition to the normal state.

\begin{figure}
	\includegraphics[width=.5\textwidth]{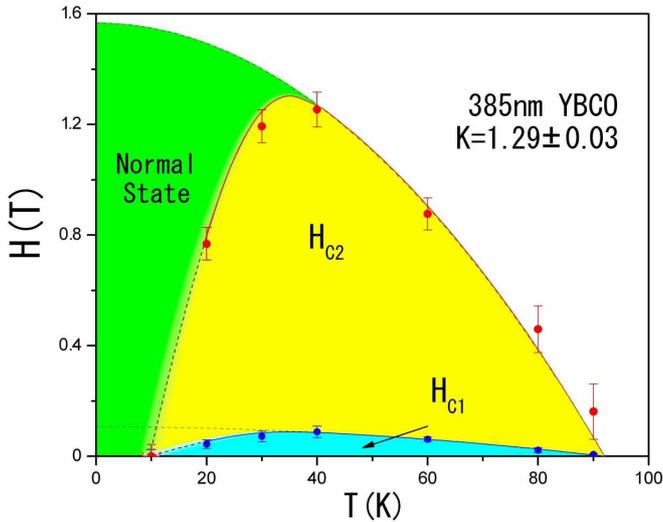}
	\caption{(Color on-line) Upper and lower critical fields of the 385 nm YBCO particles. T$ _{c1} $ = 92 $\pm$0.5 K and  T$ _{c2} $ $ \approx $ 10 $\pm$3 K.}
	\label{fig:6}       
\end{figure}

\textit{M(H)} measurements at different temperatures were made in the 385 nm YBCO particles. These particles exhibit ferromagnetism from 2 to 10 K, Figs. 5\textit{a} and \textit{b}. The presence of a large paramagnetic contribution in addition to the superconducting response is observed after 10 K. Hysteresis loops at 2, 20, and 300 K show the evolution of the magnetic behavior. A small coercive field $\approx$ 180 Oe at 2 K confirms the presence of the ferromagnetic order while at 20 K the diamagnetic contribution prevails in the isotherm. As a comparison, a high-temperature hysteresis loop of the pressed and sintered YBCO particles is shown. Both 385 nm particles and sintered powder show paramagnetism at 300 K. On the other hand, Fig. 6 shows the upper and lower critical fields as a function of the temperature. The \textit{k} parameter was found as $\sim $  1.29 $\pm$.03, which is drastically reduced in contrast with that reported by Riseman et al. in YBCO single crystals \cite{Riseman}.

Finally, three important points to note from this study are; first, the T$ _{c1}$ did not change with the particle size. Hence, it is discarded that the particles present oxygen deficit by a non-chemical equilibrium state as was reported by Hasanain and Zhonghua et al. \cite{Hasanain} and \cite{Zhonghua}. It is important to point out that normally, a change in the stoichiometry of a ceramic superconductor implies a change in the T$ _{c} $ since this superconducting characteristic is highly dependent on the chemical composition. Second, the strong influence of the vortex flux dynamics induced by sample microstructure results in a wide superconducting transition in $\chi(T)$ measurements which contrasts with the narrow transition observed in single crystals or textured samples \cite{Jin} \cite{Dersch} and \cite{Schneemeyer}. Third, the absence of measurable ferromagnetism in bulk polycrystalline samples is attributed to the masking caused by the strongest diamagnetic component, where the ferromagnetism will be unlooked unless diamagnetic contribution decreases.

\section{Conclusions}

A magnetic study in YBCO microstructured particles is presented. It is proposed that the reentrant superconducting effect is caused by the coexistence between bulk superconductivity and surface ferromagnetism where the enhanced magneto-electric coupling induced by broken spatial-inversion symmetry in layers near structural defects plays a very important role on surface ferromagnetism. The study considers that grain boundaries inside the particles act as additional surface regions, hence the increase of surface/volume ratio weakens the superconductivity and enhances the ferromagnetism simultaneously. At low temperatures and small particle sizes, the electromagnetic interaction between both phenomena tends to suppress the superconductivity, observing the transition to the normal state.

This study particularly shows the magnetic properties as a function of particle size. Upper and lower critical fields and the transition temperatures were evaluated in the 385 nm particles. These particles showed the reentrant superconducting effect. The Ginzburg-Landau parameter and the critical temperatures are \textit{k} $\sim $ 1.29 $\pm$.03,  T$ _{c1} $ = 92 $\pm$0.5 K, and  T$ _{c2} $ $ \approx $ 10 $\pm$3 K.

\begin{acknowledgements}
This work was supported by DGAPA-UNAM project IT100217 and CONACyT project 254280. Rodolfo E. L\'{o}pez-Romero thanks to CONACyT-M\'{e}xico for scholarship, Dr. F. Ascencio and R. Herrera for their help in DLS technique, Dr. F. Morales for their significant comments and Ana K. Bobadilla for helium supplies. We thanks to A. L\'{o}pez and A. Pompa Garc\'{i}a for technical support.
\end{acknowledgements}

\bibliographystyle{spphys}       
\bibliography{YBCO_article}   

\begin{thebibliography}{10}
\providecommand{\url}[1]{{#1}}
\providecommand{\urlprefix}{URL }
\expandafter\ifx\csname urlstyle\endcsname\relax
  \providecommand{\doi}[1]{DOI \discretionary{}{}{}#1}\else
  \providecommand{\doi}{DOI \discretionary{}{}{}\begingroup
  \urlstyle{rm}\Url}\fi

\bibitem{Rogalla}
H.~Rogalla, P.H. Kes, \emph{100 Years of Superconductivity} (CRC Press Taylor
  and Francis Group, New York, 2012)

\bibitem{Riblet}
G.~Riblet, K.~Winzer, Solid State Communications \textbf{9}, 1663 (1971)

\bibitem{Fertig}
W.~Fertig, D.~Johnston, L.~DeLong, R.~McCallum, M.~Maple, B.~Matthias, Physical
  Review Letters \textbf{38}, 987 (1977)

\bibitem{Ishikawa}
M.~Ishikawa, {\O}.~Fischer, Solid State Communications \textbf{23}, 37 (1977)

\bibitem{Zdravkov}
V.~Zdravkov, A.~Sidorenko, G.~Obermeier, S.~Gsell, M.~Schreck, C.~M{\"u}ller,
  S.~Horn, R.~Tidecks, L.~Tagirov, Physical review letters \textbf{97}, 057004
  (2006)

\bibitem{Reich}
S.~Reich, G.~Leitus, R.~Popovitz-Biro, M.~Schechter, Physical review letters
  \textbf{91}, 147001 (2003)

\bibitem{HLi}
W.H. Li, C.W. Wang, C.Y. Li, C.~Hsu, C.~Yang, C.M. Wu, Physical Review B
  \textbf{77}, 094508 (2008)

\bibitem{WHLi}
W.H. Li, C.~Yang, F.~Tsao, S.~Wu, P.~Huang, M.~Chung, Y.~Yao, Physical Review B
  \textbf{72}, 214516 (2005)

\bibitem{Yuewei}
Y.~Yin, H.~Liu, L.~Xie, T.~Su, M.~Teng, X.~Li, The Journal of Physical
  Chemistry C \textbf{117}, 3028 (2013)

\bibitem{Shipra}
A.~Shipra, Gomathi, A.~Sundaresan, C.~Rao, et~al., Solid state communications
  \textbf{142}, 685 (2007)

\bibitem{Hasanain}
S.~Hasanain, N.~Akhtar, A.~Mumtaz, Journal of Nanoparticle Research
  \textbf{13}, 1953 (2011)

\bibitem{Zhonghua}
Z.~Zhu, D.~Gao, C.~Dong, G.~Yang, J.~Zhang, J.~Zhang, Z.~Shi, H.~Gao, H.~Luo,
  D.~Xue, Physical Chemistry Chemical Physics \textbf{14}, 3859 (2012)

\bibitem{Stoneham}
A.~Stoneham, L.~Smith, Journal of Physics: Condensed Matter \textbf{3}, 225
  (1991)

\bibitem{ASundaresan}
A.~Sundaresan, C.~Rao, Nano Today \textbf{4}, 96 (2009)

\bibitem{Sunda}
A.~Sundaresan, C.~Rao, Solid state communications \textbf{149}, 1197 (2009)

\bibitem{Fan}
W.~Fan, L.J. Zou, Z.~Zeng, Physica C: Superconductivity \textbf{492}, 80 (2013)

\bibitem{Coey}
J.~Coey, K.~Wongsaprom, J.~Alaria, M.~Venkatesan, Journal of Physics D: Applied
  Physics \textbf{41}, 134012 (2008)

\bibitem{Osorio}
J.~Osorio-Guill{\'e}n, S.~Lany, S.~Barabash, A.~Zunger, Physical review letters
  \textbf{96}, 107203 (2006)

\bibitem{Chaitanya}
C.D. Pemmaraju, S.~Sanvito, Physical review letters \textbf{94}, 217205 (2005)

\bibitem{Junling}
J.~Wang, \emph{Multiferroic Materials Properties, Techniques, and Applications}
  (CRC Press Taylor and Francis Group, New York, 2017)

\bibitem{Jin}
S.~Jin, T.~Tiefel, R.~Sherwood, M.~Davis, R.~Van~Dover, G.~Kammlott,
  R.~Fastnacht, H.~Keith, Applied physics letters \textbf{52}, 2074 (1988)

\bibitem{Blinov}
E.~Blinov, V.~Fleisher, H.~Huhtinen, R.~Laiho, E.~L{\"a}hderanta, P.~Paturi,
  Y.P. Stepanov, L.~Vlasenko, Superconductor Science and Technology
  \textbf{10}, 818 (1997)

\bibitem{AManthiram}
A.~Manthiram, J.B. Goodenough, Nature \textbf{329}, 701 (1987)

\bibitem{J.Gotor}
F.~Gotor, P.~Simon, N.~Pellerin, P.~Odier, P.~Monod, Journal of Physics and
  Chemistry of Solids \textbf{58}, 1469 (1997)

\bibitem{Maple}
M.~Maple, W.~Fertig, A.~Mota, L.~DeLong, D.~Wohlleben, R.~Fitzgerald, Solid
  State Communications \textbf{11}, 829 (1972)

\bibitem{Wahle}
J.~Wahle, N.~Bl{\"u}mer, J.~Schlipf, K.~Held, D.~Vollhardt, Physical review B
  \textbf{58}, 12749 (1998)

\bibitem{Riseman}
T.~Riseman, J.~Brewer, K.~Chow, W.~Hardy, R.~Kiefl, S.~Kreitzman, R.~Liang,
  W.~MacFarlane, P.~Mendels, G.~Morris, et~al., Physical Review B \textbf{52},
  10569 (1995)

\bibitem{Dersch}
H.~Dersch, G.~Blatter, Physical Review B \textbf{38}, 11391 (1988)

\bibitem{Schneemeyer}
L.~Schneemeyer, J.~Waszczak, T.~Siegrist, R.~Van~Dover, L.~Rupp, B.~Batlogg,
  R.J. Cava, D.~Murphy, Nature \textbf{328}, 601 (1987)

\end{thebibliography}

\end{document}